\documentclass[8.5pt,twoside,twocolumn,amsmath,amssymb]{article}
\usepackage[super,sort&compress,comma]{natbib}
\usepackage[version=3]{mhchem}
\usepackage{times,mathptmx}
\usepackage{sectsty}
\usepackage{balance}

\usepackage{graphicx} 
\usepackage{lastpage}
\usepackage[format=plain,singlelinecheck=false,font=small,labelfont=bf,labelsep=space]{caption}
\usepackage{fancyhdr}
\usepackage{dcolumn}
\usepackage{bm}
\usepackage{amsmath}
\usepackage{bbding}
\usepackage{verbatim}

\newcommand{\etalia}{{\it et al.~}}
\newcommand{\la}{\left\langle}
\newcommand{\ra}{\right\rangle}

\newcommand{\PRL}{Phys.~Rev.~Lett.~}

\begin{document}

\thispagestyle{plain}
\fancypagestyle{plain}{
\fancyhead[L]{\includegraphics[height=8pt]{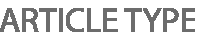}}
\fancyhead[C]{\hspace{-1cm}\includegraphics[height=20pt]{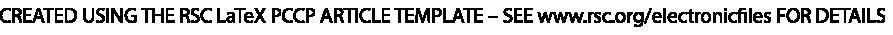}}
\fancyhead[R]{\includegraphics[height=10pt]{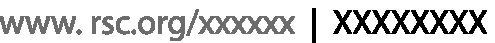}\vspace{-0.2cm}}
\renewcommand{\headrulewidth}{1pt}}
\renewcommand{\thefootnote}{\fnsymbol{footnote}}
\renewcommand\footnoterule{\vspace*{1pt}%
\hrule width 3.4in height 0.4pt \vspace*{5pt}}
\setcounter{secnumdepth}{5}

\makeatletter
\def\subsubsection{\@startsection{subsubsection}{3}{10pt}{-1.25ex plus -1ex minus -.1ex}{0ex plus 0ex}{\normalsize\bf}}
\def\paragraph{\@startsection{paragraph}{4}{10pt}{-1.25ex plus -1ex minus -.1ex}{0ex plus 0ex}{\normalsize\textit}}
\renewcommand\@biblabel[1]{#1}
\renewcommand\@makefntext[1]%
{\noindent\makebox[0pt][r]{\@thefnmark\,}#1}
\makeatother
\renewcommand{\figurename}{\small{Fig.}~}
\sectionfont{\large}
\subsectionfont{\normalsize}

\fancyfoot{}
\fancyfoot[LO,RE]{\vspace{-7pt}\includegraphics[height=9pt]{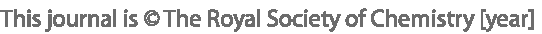}}
\fancyfoot[CO]{\vspace{-7.2pt}\hspace{12.2cm}\includegraphics{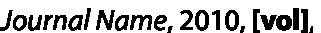}}
\fancyfoot[CE]{\vspace{-7.5pt}\hspace{-13.5cm}\includegraphics{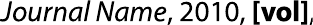}}
\fancyfoot[RO]{\footnotesize{\sffamily{1--\pageref{LastPage} ~\textbar  \hspace{2pt}\thepage}}}
\fancyfoot[LE]{\footnotesize{\sffamily{\thepage~\textbar\hspace{3.45cm} 1--\pageref{LastPage}}}}
\fancyhead{}
\renewcommand{\headrulewidth}{1pt}
\renewcommand{\footrulewidth}{1pt}
\setlength{\arrayrulewidth}{1pt}
\setlength{\columnsep}{6.5mm}
\setlength\bibsep{1pt}

\twocolumn[
  \begin{@twocolumnfalse}
\noindent\LARGE{\textbf{Swelling, Structure, and Phase Stability of Compressible Microgels}}
\vspace{0.6cm}

\noindent\large{\textbf{Matthew Urich and Alan R. Denton\textit{$^{\ast}$}
}}\vspace{0.5cm}

\noindent\textit{\small{\textbf{Received 8th September 2016, Accepted 17th October 2016\newline
First published on the web 17th October 2016}}}

\noindent \textbf{\small{DOI: 10.1039/b000000x}}
\vspace{0.6cm}

\noindent \normalsize{
Microgels are soft colloidal particles that, when dispersed in a solvent, swell and
deswell in response to changes in environmental conditions, such as temperature,
concentration, and $p$H.  Using Monte Carlo simulation, we model bulk suspensions
of microgels that interact via Hertzian elastic interparticle forces and can expand
or contract via trial moves that allow particles to change size in accordance with
the Flory-Rehner free energy of cross-linked polymer gels.  We monitor the influence
of particle compressibility, size fluctuations, and concentration on bulk structural
and thermal properties by computing particle swelling ratios, radial distribution
functions, static structure factors, osmotic pressures, and freezing densities.
For microgels in the nanoscale size range, particle compressibility and associated
size fluctuations suppress crystallization, shifting the freezing transition to a
higher density than for the hard-sphere fluid.  As densities increase beyond
close packing, microgels progressively deswell, while their intrinsic size distribution
grows increasingly polydisperse.
}
\vspace{0.6cm}
 \end{@twocolumnfalse}
  ]

\footnotetext{\textit{Department of Physics, North Dakota State University, 
Fargo, ND 58108-6050, USA.  E-mail: alan.denton@ndsu.edu}}

%
%
%
%
%
%
%
%
%
%
%
%

\section{Introduction}

Microgels are soft, compressible colloidal particles, typically composed of cross-linked 
polymer networks that, when dispersed in a solvent, can swell significantly in size and 
can respond sensitively to environmental changes.  Equilibrium particle sizes are 
determined both by the elasticity of the gel network and by temperature, particle 
concentration, and solvent quality.\cite{baker1949,pelton1986,pelton2000,saunders2009}
Ionic microgels, which acquire charge via dissociation of counterions, further respond to 
changes in $p$H and salt concentration.  Tunable particle size results in unusual materials 
properties, with practical applications to filtration, rheology, and drug delivery.\cite{
HydrogelBook2012,MicrogelBook2011,lyon-nieves-AnnuRevPhysChem2012,yunker-yodh-review2014}

Over the past two decades, numerous experimental and modeling studies have characterized 
the elastic properties of single microgel particles\cite{
cloitre-leibler1999,
cloitre-leibler2003,
tan2004,
nieves-macromol2000,
nieves-jcp2003,
nieves-macromol2009,
hellweg2010,
weitz-sm2012,
weitz-jcp2012,
ciamarra2013,
nieves-sm2011,
nieves-sm2012,
nieves-bulk-shear-pre2011,
nieves-bulk-pre2011,
dufresne2009,
nieves-prl2015}
and the equilibrium and dynamical behavior of bulk suspensions.\cite{
weitz-prl1995,
groehn2000,
levin2002,
nieves-jcp2005,
winkler-gompper2012,
winkler-gompper2014,
li-chen2014,
egorov-likos2013,
colla-likos2014,
colla-likos2015,
stellbrink-likos-nanoscale2015,
stellbrink-likos-prl2015}
Connections between single-particle properties, such as swelling ratio, and bulk properties, 
such as osmotic pressure, thermodynamic phase behavior, pair structure, and viscosity, have been 
probed experimentally by static and dynamic light scattering, small-angle neutron scattering, 
confocal microscopy, and osmometry.\cite{mohanty-richtering2008,richtering2008,
lyon2007,weitz-pre2012,schurtenberger-ZPC2012,schurtenberger-SM2012,schurtenberger2013,
nieves-pre2013,schurtenberger2014}
While the swelling/deswelling behavior of microgels has been extensively studied, the 
full implications of elasticity and compressibility of these soft colloids for bulk 
suspension properties remain only partially understood.  

Previous modeling studies have examined the influence of elastic interparticle 
interactions on structure and phase behavior.  In an extensive simulation survey,
Pami\`es \etalia\cite{frenkel2009} mapped out the thermodynamic phase diagram of
a model of elastic, but incompressible, spheres interacting via a Hertz pair 
potential.\cite{landau-lifshitz1986}  For a related model of ionic microgels, 
one of us and coworkers\cite{hedrick-chung-denton2015} recently applied 
molecular dynamics simulation and thermodynamic perturbation theory to compute the 
osmotic pressure and static structure factor of a suspension of Hertzian spheres that 
interact also electrostatically, via an effective Yukawa pair potential.\cite{denton2000}
At the single-particle level, numerous studies have established that the classic 
Flory-Rehner theory of swelling of cross-linked polymer networks,\cite{flory1953}
though originally developed for macroscopic gels, provides a reasonable description 
also of the elastic properties of microgel particles.\cite{weitz-jcp2012,
ciamarra2013,
colla-likos2014,
colla-likos2015,
nieves-sm2012,
weitz-sm2012,
nieves-bulk-shear-pre2011,
nieves-sm2011,
nieves-bulk-pre2011,
nieves-macromol2009,
nieves-jcp2003,
nieves-macromol2000,
pelton2000,
hellweg2010,
saunders2009,
moncho-jorda-dzubiella2016}
However, the implications of particle compressibility and intrinsic size polydispersity 
for bulk properties of microgel suspensions have not been fully explored.
The purpose of this paper is to analyze the combined influences of particle elasticity,
compressibility, and associated size fluctuations on bulk thermal and structural properties 
of microgel suspensions.  

The remainder of the paper is organized as follows.  In Sec.~\ref{model},
we describe a model of microgels as compressible spheres, whose swelling
is governed by a single-particle free energy derived from the Flory-Rehner
theory of cross-linked polymer networks, and whose interparticle interactions
are represented by a Hertz elastic pair potential.  In Sec.~\ref{methods}, 
we outline Monte Carlo simulation methods by which we modeled bulk suspensions 
of compressible, size-fluctuating microgels. 
Section~\ref{results} presents numerical results for thermodynamic properties,
specifically osmotic pressure and liquid-solid phase behavior, and structural 
properties, including particle volume fraction, radial distribution function, 
and static structure factor.  Finally, in Sec.~\ref{conclusions}, we summarize 
and conclude.

\section{Model}\label{model}

\subsection{Swelling of Microgel Particles}\label{particles}
\begin{figure}[t!]
\includegraphics[width=0.9\columnwidth,angle=0]{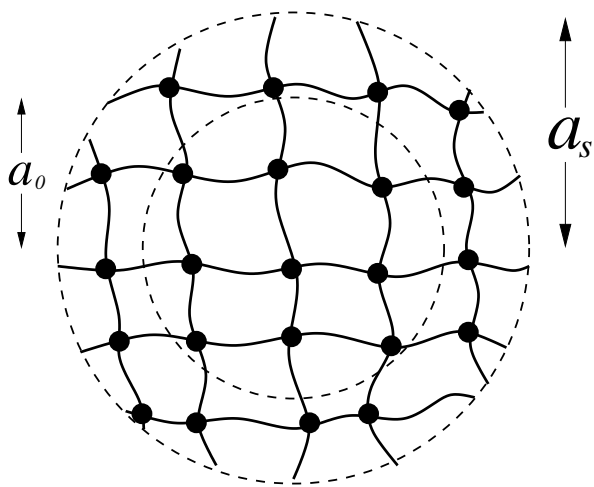}
\vspace*{-0.2cm}
\caption{
Schematic drawing of a microgel of dry radius $a_0$ swollen by solvent to a radius $a_s$. 
}\label{fig1}
\end{figure}

We model a microgel as a spherical particle of dry (collapsed, unswollen) radius $a_0$ 
and swollen radius $a$, consisting of a cross-linked network of $N_m$ monomers, with 
a uniform distribution of cross-linkers that divide the network into $N_{\rm ch}$ 
distinct chains (see Fig.~\ref{fig1}).  Although idealized, this simple model 
provides a reference that can be generalized to heterogeneous microgels with 
nonuniform distribution of cross-linkers, such as core-shell\cite{stieger2004,
nieves-sm2011,weitz-jcp2012,schurtenberger-ZPC2012,ciamarra2013,
moncho-jorda-anta2013} or 
%
hollow\cite{quesada-perez2013,quesada-perez-moncho-jorda-anta2015,potemkin2015} microgels.  
To highlight the interplay between intra- and inter-particle elasticity, without the 
added complexity of electrostatic interactions, we consider here only nonionic microgels.

For the uniform-sphere model, with particle size swelling ratio $\alpha\equiv a/a_0$, 
the Flory-Rehner theory of polymer networks combines mixing entropy, polymer-solvent interactions, 
and elastic free energy to predict the total Helmholtz free energy of the network:\cite{flory1953}
\begin{eqnarray}
\beta F(\alpha)&=&N_m\left[(\alpha^3-1)\ln\left(1-\alpha^{-3}\right) 
+\chi\left(1-\alpha^{-3}\right)\right] 
\nonumber\\[1ex]
&+&\frac{3}{2}N_{\rm ch}\left(\alpha^2-\ln\alpha-1\right)~,
\label{FloryF}
\end{eqnarray}
where $\beta\equiv 1/(k_BT)$ at temperature $T$ and $\chi$ is the polymer-solvent interaction 
(solvency) parameter.  
Swelling of a microgel particle results from stretching of polymer coils, subject to 
the constraints of a cross-linked network.\cite{likos01}
Of the two terms in square brackets in Eq.~(\ref{FloryF}), the first accounts for 
the entropy of mixing of the microgel with solvent molecules, setting to zero 
the number of individual polymer chains in the network structure.  The second term 
represents a mean-field approximation for the interaction between polymer monomers 
and solvent molecules, which entirely neglects interparticle correlations.  
The last term in Eq.~(\ref{FloryF}) describes the elastic free energy associated with 
stretching the microgel, assuming isotropic deformation, neglecting any change in 
internal energy of the network, and modeling the polymer chains as Gaussian coils.
The assumption of a purely entropic elastic free energy ignores enthalpic contributions 
stemming from structural changes in surrounding solvent upon stretching a polymer coil,
while the Gaussian coil approximation is valid only for swelling ratios not exceeding
the polymer contour length.\cite{potemkin2015}
Given the Flory-Rehner free energy [Eq.~(\ref{FloryF})], the size of a single, 
isolated microgel (i.e., in a dilute solution) fluctuates according to a 
probability distribution, 
\begin{equation}
P_0(\alpha)~\propto~\exp[-\beta F(\alpha)]~.
\label{FloryP}
\end{equation}
In thermal equilibrium, an isolated, swollen microgel in a dilute solution has a most probable 
radius $a_s$, corresponding to the maximum of this distribution (i.e., minimum free energy).
A suspension of particles thus has a fluctuating particle size distribution, i.e., dynamical 
size polydispersity, governed by Eq.~(\ref{FloryP}).

\subsection{Microgel Pair Interactions}\label{interactions}
While swelling of isolated microgels in dilute solutions is governed only by intraparticle
interactions, modeled by Eq.~(\ref{FloryF}), swelling in concentrated solutions is 
influenced also by interparticle interactions.  To account for this additional influence,
we model the repulsion between a pair of microgel particles,
of instantaneous radii $a_i$ and $a_j$ at center-to-center separation $r$,
via a Hertz effective pair potential,\cite{landau-lifshitz1986}
\begin{equation}
v_H(r)=\left\{ \begin{array} 
{l@{\quad\quad}l}
\epsilon\left(1-\frac{\displaystyle r}{\displaystyle a_i+a_j}\right)^{5/2}~, 
& r<a_i+a_j \\[1ex]
0~, 
& r\ge a_i+a_j~. \end{array} \right.
\label{Hertz}
\end{equation}
The total internal energy associated with pair interactions is then given by
\begin{equation}
U=\sum_{i<j=1}^N\, v_H(r_{ij})~,
\label{U}
\end{equation}
where $r_{ij}$ is the distance between the centers of particles $i$ and $j$.
For particles of average volume $v=(4\pi/3)\la a^3\ra$, with angular brackets denoting 
an ensemble average over configurations, the Hertz pair potential amplitude is 
\begin{equation}
\epsilon=\frac{4Y v}{5\pi(1-\nu^2)}~,
\label{epsilon}
\end{equation}
which depends on the elastic properties of the gel through Young's modulus $Y$ 
and the Poisson ratio $\nu$.\cite{landau-lifshitz1986}  Scaling theory of 
polymer gels in good solvents\cite{deGennes1979} predicts that the bulk modulus 
scales linearly with temperature and density of cross-linkers (or chain number): 
$Y\sim TN_{\rm ch}/v$.  The denser the cross-links, the stiffer the gel.
It follows that the reduced amplitude, $\epsilon^*\equiv\beta\epsilon$, is
essentially independent of temperature and particle volume, neglecting 
dependence of $\nu$ on $\alpha$, and scales linearly with $N_{\rm ch}$.

The elastic properties of bulk, water-swollen hydrogels have been measured using 
scanning force microscopy.\cite{matzelle2003}  For poly(N-isopropylacrylamide) 
(PNIPAM) hydrogel with 0.25 mol \% cross-linker, Young's modulus was determined 
as 0.33 kPa in the fully swollen state at 10 $^{\circ}$C and 13.9 kPa in the 
collapsed state at 40 $^{\circ}$C.  In contrast, poly(acrylamide) (PA) hydrogels 
with cross-linker between 1 and 5 mol \% have Young's moduli measured in the range 
from 90 to 465 kPa at 20 $^{\circ}$C.  Typical Poisson ratios for these gels are 
$\nu\simeq 0.5$.  In Sec.~\ref{results}, we appeal to such measurements to select 
realistic pair potential parameters for our simulations.

\subsection{Suspensions of Swollen Microgels}\label{suspensions}
At constant $T$, the equilibrium thermodynamic state of a suspension of $N$ 
particles in a volume $V$ depends on the average number density, $n=N/V$.
For a suspension of microgels, the dry volume fraction, $\phi_0=(4\pi/3)n a_0^3$, 
is defined as the fraction of the total volume occupied by the particles in their 
dry state.  Correspondingly, the swollen volume fraction is defined as the ratio 
of the most probable volume of a swollen particle to the volume per particle:
$\phi=(4\pi/3)n\la a^3\ra=\phi_0\la\alpha^3\ra$.  In suspensions of highly swollen 
particles, $\phi$ can substantially exceed $\phi_0$.  For reference, we also define 
the {\it generalized} volume fraction\cite{nieves-bulk-pre2011,ciamarra2013} as 
the ratio of the volume of a particle of most probable size in the dilute limit 
to the volume per particle: $\zeta=(4\pi/3)na_s^3=\phi_0\alpha_s^3$.  Note that 
$\zeta\ge\phi$ (since $a_s\ge\la a\ra$) and that $\phi$ and $\zeta$ have no upper bounds.
In particular, in concentrated suspensions beyond close packing, it is possible that 
$\zeta>1$ and $\phi>1$.  At such high concentrations, crowded microgel particles 
may not only compress in size, but may also distort in shape.  The presently studied
model allows the former, but not the latter, response to crowding, although the
Hertz potential may be interpreted as allowing for faceted deformations.  In a 
more refined model, the microgels could be represented as 
elastic spheres\cite{riest2015,cloitre-bonnecaze2010,cloitre-bonnacaze2011} 
or ellipsoids\cite{miller2010,lim-denton2014,lim-denton-JCP2016,lim-denton-SM2016} 
that can compress along three axes with an associated deformation energy.

\section{Computational Methods}\label{methods}
To study the influence of particle compressibility and fluctuating size polydispersity 
on thermal and structural properties of bulk microgel suspensions, we performed a series 
of constant-$NVT$ Monte Carlo (MC) simulations of systems of particles modeled by the 
Flory-Rehner size distribution [Eq.~(\ref{FloryP})] and the Hertz pair potential 
[Eq.~(\ref{Hertz})], as described in Sec.~\ref{model}.  For $N$ particles in a cubic box 
of fixed volume $V$ with periodic boundary conditions at fixed temperature $T$,
we made trial moves consisting of combined particle displacements and size changes.
Following the standard Metropolis algorithm,\cite{frenkel-smit2001,binder-heermann2010}
a trial displacement and change in swelling ratio, from $\alpha$ to $\alpha'$,
was accepted with probability 
\begin{equation}
{\cal P}_{\rm acc} = \min\left\{\exp[-\beta(\Delta U+\Delta F)],~1\right\}~,
\label{Pacc}
\end{equation}
where $\Delta U$ is the change in internal energy [Eq.~(\ref{U})] associated with
interparticle interactions and $\Delta F=F(\alpha')-F(\alpha)$ is the change in free energy 
[Eq.~(\ref{FloryF})] associated with swelling.  At equilibrium, the particles adopt a 
size distribution $P(\alpha; \phi_0)$ that depends on the dry particle volume fraction $\phi_0$ 
and minimizes the total free energy of the system.  
Since the particles modeled here are compressible and polydisperse, the generalized 
volume fraction can exceed close packing of monodisperse hard spheres ($\phi\simeq 0.74$).
Such high concentrations create a strong interplay between the Hertz elastic 
energy and the Flory-Rehner swelling free energy.

To guide analysis of phase behavior, we computed structural properties that probe 
interparticle pair correlations and thermodynamic properties that govern phase stability.
First, we determined the radial distribution function $g(r)$ by standard means, histogramming 
into radial bins the radial separation $r$ of particle pairs in each configuration and 
averaging over configurations.  Second, we computed the static structure factor $S(q)$, 
proportional to the Fourier transform of $g(r)$ and to the intensity of scattered 
radiation at scattered wave vector magnitude $q$.  For a system of $N$ particles, 
the orientationally averaged two-particle static structure factor is defined as
\begin{equation}
S(q)=\frac{2}{N}\sum_{i<j=1}^{N}\la\frac{\sin(q r_{ij})}{q r_{ij}}\ra+1~,
\label{Sq}
\end{equation}
where angular brackets again denote an ensemble average over particle configurations
-- the same configurations as used to compute $g(r)$.
Finally, we computed the osmotic pressure of the system from the virial theorem 
\begin{equation}
\beta \pi/n=1+\frac{3}{Nk_BT}\sum_{i<j=1}^N\, \la r_{ij}f(r_{ij})\ra~,
\label{pressure}
\end{equation}
where $f(r)=-v_H'(r)$ is the Hertz pair force.

\begin{figure}[t!]
\includegraphics[width=0.9\columnwidth,angle=0]{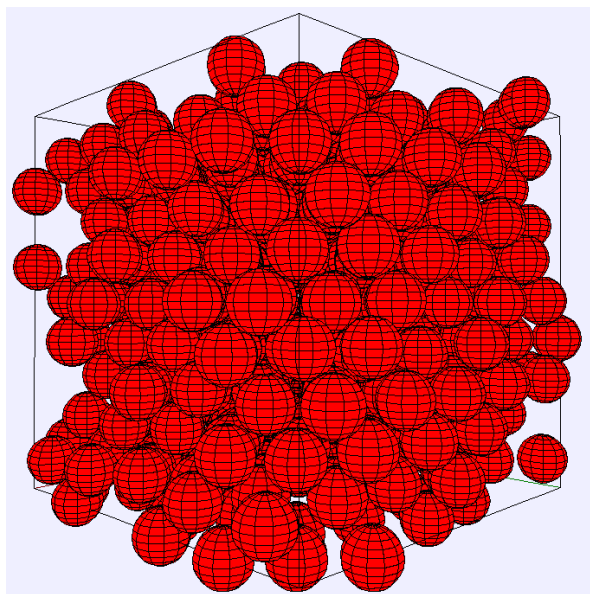}
\vspace*{-0.2cm}
\caption{
Typical snapshot from an MC simulation of a suspension of compressible, spherical 
microgel particles in a cubic box with periodic boundary conditions.
}\label{fig2}
\end{figure}
\begin{figure}[t!]
\includegraphics[width=0.9\columnwidth,angle=0]{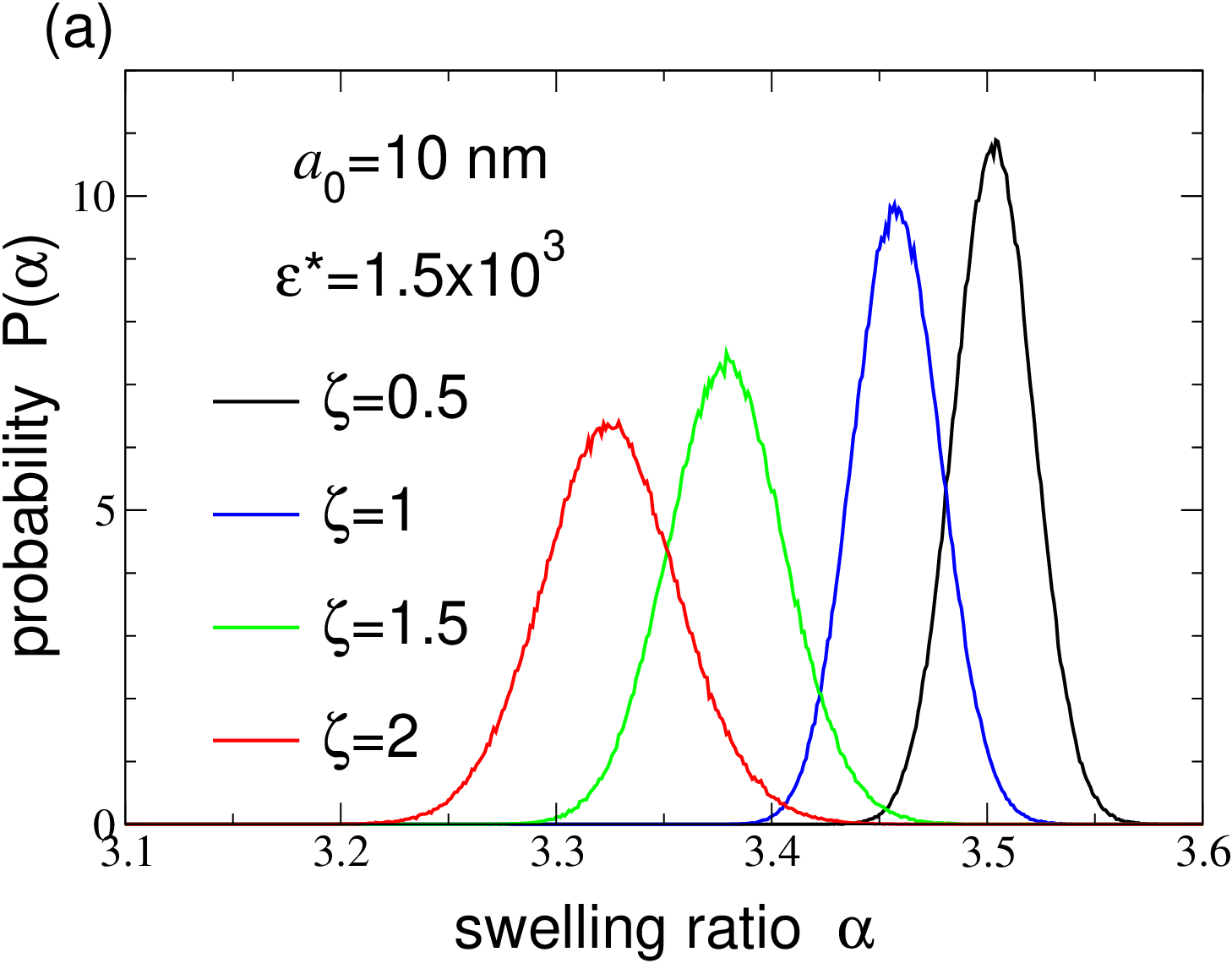}
\includegraphics[width=0.9\columnwidth,angle=0]{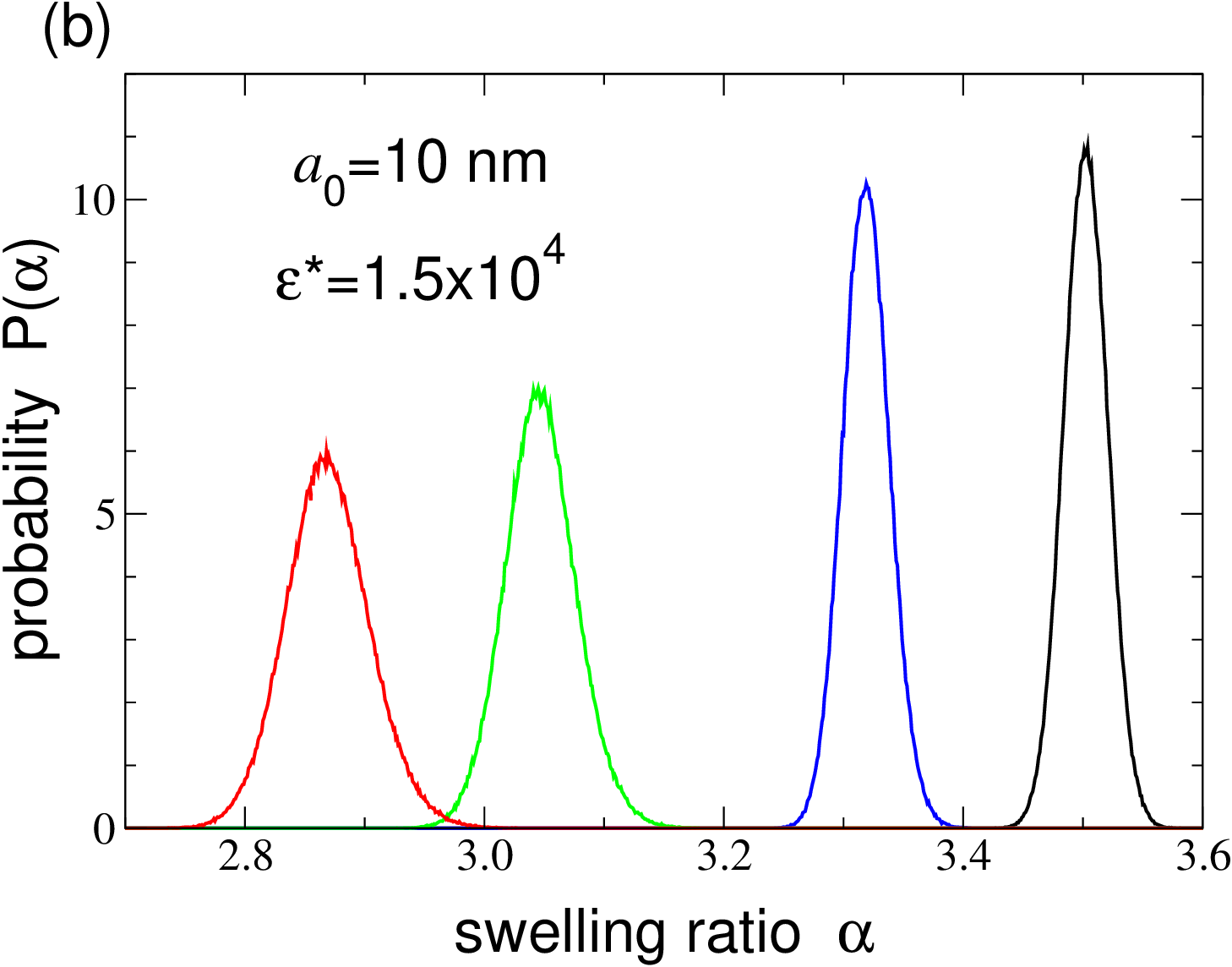}
\vspace*{-0.2cm}
\caption{
Normalized probability distribution $P(\alpha)$ of swelling ratio $\alpha$ of compressible microgels 
of dry radius $a_0=10$ nm, composed of $N_m=2\times 10^5$ monomers with $N_{\rm ch}=200$ 
chains in a solvent with Flory solvency parameter $\chi=0$ at generalized volume fractions 
$\zeta=0.5$, 1, 1.5, 2 (right to left).  The particles interact via a Hertz pair potential 
with reduced amplitude (a) $\epsilon^*=1.5\times 10^3$ and (b) $\epsilon^*=1.5\times 10^4$.  
With increasing concentration ($\zeta$), particles are increasingly compressed, as reflected 
by shift in distribution toward smaller values of $\alpha$. 
}\label{fig3}
\end{figure}
\begin{figure}[t!]
\includegraphics[width=0.9\columnwidth,angle=0]{alpha.a100.B1M.eps}
\includegraphics[width=0.9\columnwidth,angle=0]{alpha.a100.B1e7.eps}
\vspace*{-0.2cm}
\caption{
Normalized probability distribution $P(\alpha)$ of swelling ratio $\alpha$ of compressible microgels 
of dry radius $a_0=100$ nm, composed of $N_m=2\times 10^8$ monomers with $N_{\rm ch}=2\times 10^5$ 
chains in a solvent with Flory solvency parameter $\chi=0$ at generalized volume fractions 
$\zeta=1$, 1.5, 2 (right to left).  The particles interact via a Hertz pair potential 
with reduced amplitude (a) $\epsilon^*=10^6$ and (b) $\epsilon^*=10^7$, where the 
$\zeta=1$ distribution (off scale to right) is extremely narrow and sharply peaked.
Note the change in scale from Fig.~\ref{fig3}.
}\label{fig4}
\end{figure}
\begin{figure}[t!]
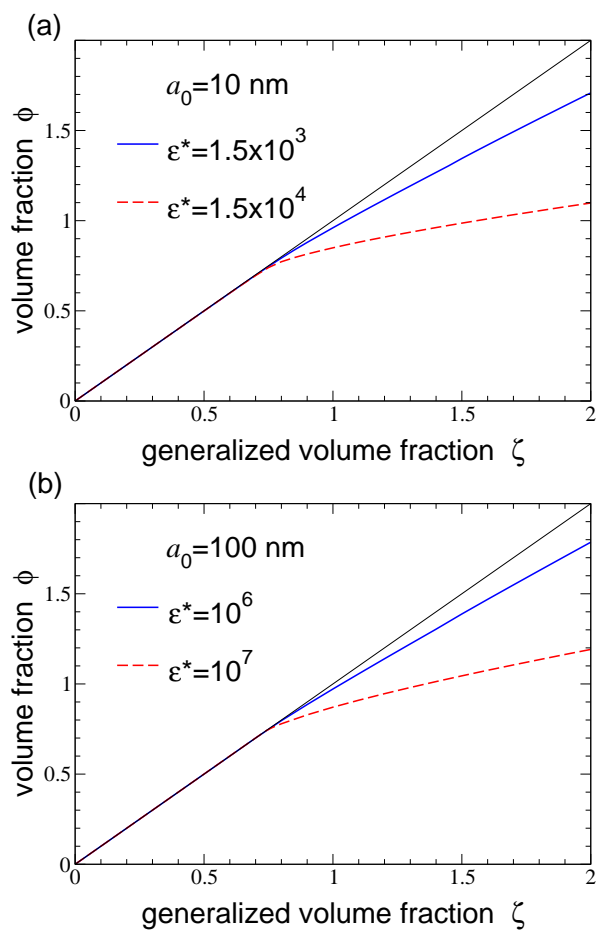

\includegraphics[width=0.9\columnwidth,angle=0]{zeta.a10.eps}
\includegraphics[width=0.9\columnwidth,angle=0]{zeta.a100.eps}
\vspace*{-0.2cm}
\caption{
Volume fraction $\phi$ vs.~generalized volume fraction $\zeta$ for 
suspensions of compressible microgels with $\chi=0$ and
(a) $a_0=10$ nm, $a_s=35.055$ nm, $N_m=2\times 10^5$, $N_{\rm ch}=200$, and
(b) $a_0=100$ nm, $a_s=350.55$ nm, $N_m=2\times 10^8$, $N_{\rm ch}=2\times 10^5$,
for various values of Hertz pair potential reduced amplitude $\epsilon^*$.
The line $\phi=\zeta$ is drawn for reference.
With increasing $\zeta$, self-crowding compresses particles beyond close packing, 
causing $\phi$ to increase more slowly than $\zeta$.
}\label{fig5}
\end{figure}
\begin{figure}
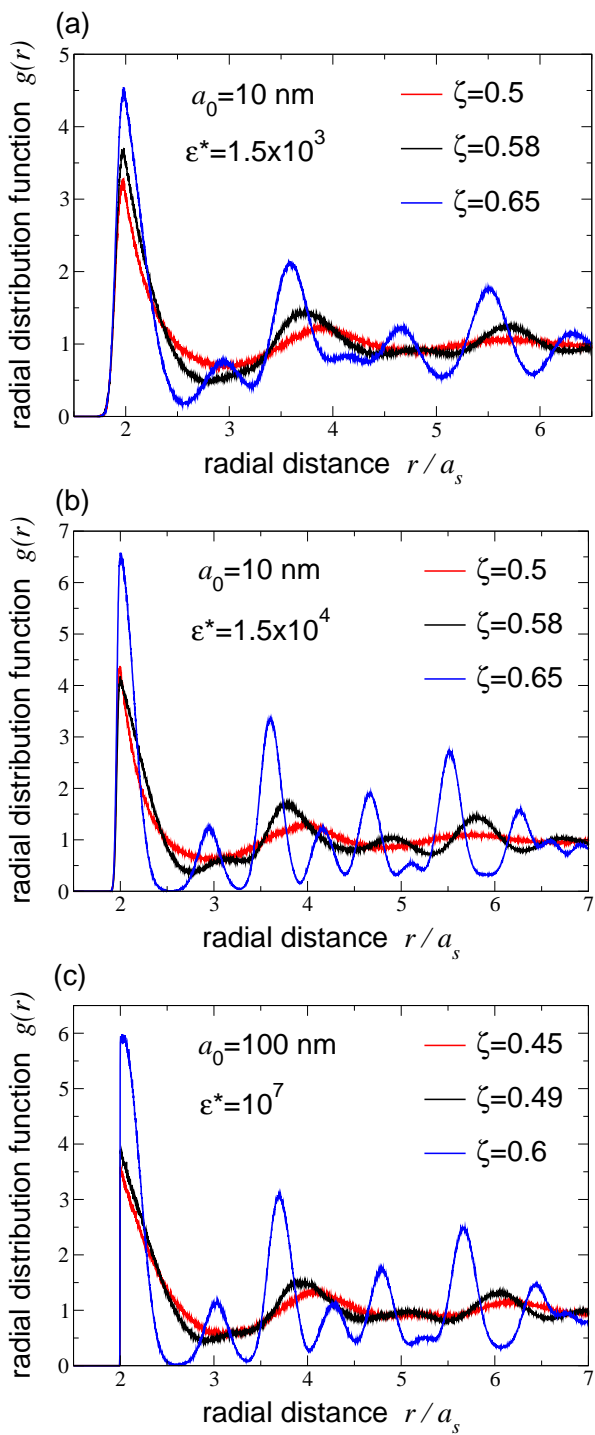

\includegraphics[width=0.9\columnwidth,angle=0]{rdf.a10.B1500.eps}
\includegraphics[width=0.9\columnwidth,angle=0]{rdf.a10.B15000.eps}
\includegraphics[width=0.9\columnwidth,angle=0]{rdf.a100.B1e7.eps}
\vspace*{-0.2cm}
\caption{
Radial distribution functions $g(r)$ vs.~radial distance $r$ (in units of uncrowded 
swollen radius $a_s$) of microgel suspensions, of generalized volume fractions $\zeta$
straddling freezing transition (black curves), with same particle parameters as in 
Fig.~\ref{fig5} and Hertz pair potential reduced amplitudes 
(a) $\epsilon^*=1.5\times 10^3$, (b) $\epsilon^*=1.5\times 10^4$, and (c) $\epsilon^*=10^7$.
Red and blue curves correspond to fluid and solid phases, respectively.
}\label{fig6}
\end{figure}
\begin{figure}
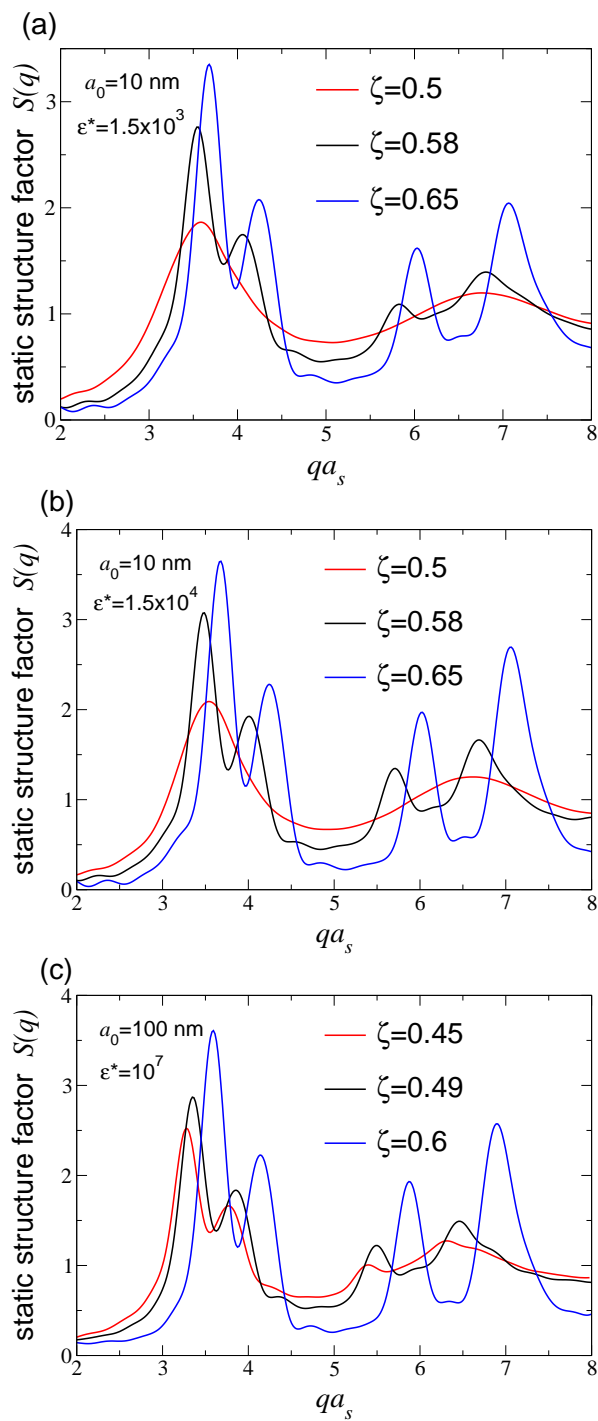

\includegraphics[width=0.9\columnwidth,angle=0]{sq.a10.B1500.eps}
\includegraphics[width=0.9\columnwidth,angle=0]{sq.a10.B15000.eps}
\includegraphics[width=0.9\columnwidth,angle=0]{sq.a100.B1e7.eps}
\vspace*{-0.2cm}
\caption{
Static structure factors $S(q)$ vs.~scattered wave vector magnitude $q$
[from Eq.~(\ref{Sq})] corresponding to radial distribution functions in Fig.~\ref{fig6}.
Freezing occurs when main peak height $S(q_{\rm max})>2.8$ (black curves).
Red and blue curves correspond to fluid and solid phases, respectively.
}\label{fig7}
\end{figure}
\begin{figure}
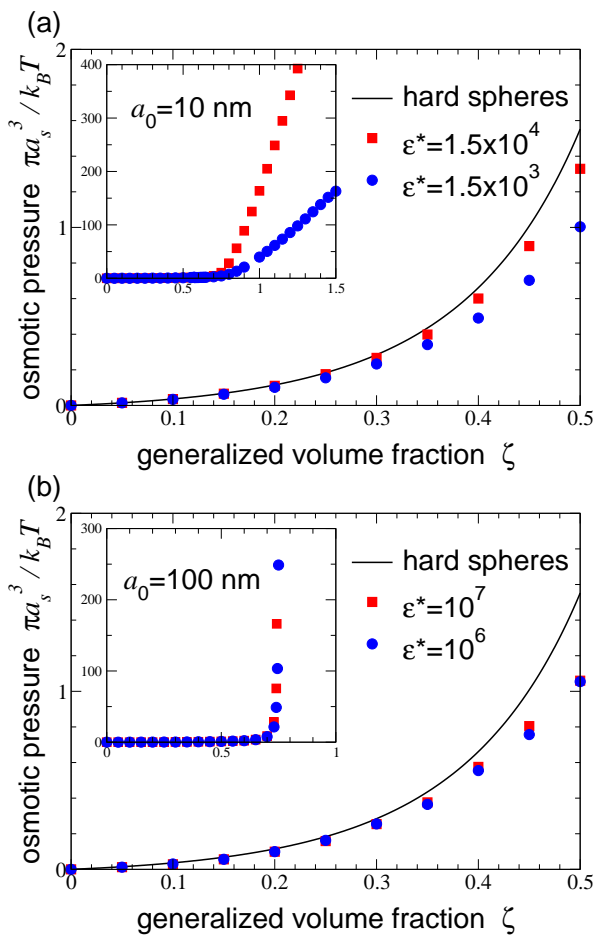

\includegraphics[width=0.9\columnwidth,angle=0]{p.a10.eps}
\includegraphics[width=0.9\columnwidth,angle=0]{p.a100.eps}
\vspace*{-0.2cm}
\caption{
Equation of state: Reduced osmotic pressure $\beta\pi a_s^3$ vs.~generalized volume fraction 
$\zeta$ [from Eq.~(\ref{pressure})] for suspensions of compressible microgels with $\chi=0$ 
and (a) $a_0=10$ nm, $a_s=35.055$ nm, $N_m=2\times 10^5$, $N_{\rm ch}=200$, and
(b) $a_0=100$ nm, $a_s=350.55$ nm, $N_m=2\times 10^8$, $N_{\rm ch}=2\times 10^5$,
for various values of Hertz pair potential reduced amplitude $\epsilon^*$.
Insets display broader ranges of $\zeta$ and $\pi$.
Pressure of hard-sphere fluid ($\epsilon^*\to\infty$) is shown for reference.
With increasing $\epsilon^*$, osmotic pressure becomes more hard-sphere-like.
}\label{fig8}
\end{figure}

\section{Results and Discussion}\label{results}
From simulations of $N=500$ particles initialized on an fcc lattice, we analyzed 
the equilibrium particle swelling ratio, structural properties, and osmotic pressure 
over a wide range of densities.  The results presented below represent statistical 
averages of particle coordinates and radii over 1000 independent configurations, 
separated by intervals of 100 MC steps (total of $10^5$ steps), following an initial 
equilibration phase of $5\times 10^4$ MC steps, where a single MC step is defined as 
one combined trial displacement and size change of every particle.  To confirm that 
the system had equilibrated, we computed the total internal energy and pressure and 
checked that both quantities had reached stable plateaus.  
A typical snapshot of the system is shown in Fig.~\ref{fig2}.  

In the low-density (dilute) regime, the initial lattice structure rapidly fell apart,
as particles freely diffused through the system, implying a stable fluid phase.
In the high-density (concentrated) regime, strongly interacting particles remained 
in the fcc lattice structure, consistent with a stable or metastable crystalline phase.
At intermediate densities, relative stability of fluid and solid phases depended on 
the interparticle interactions.  We emphasize that we did not attempt to determine 
thermodynamic phase stability, which would require more extensive simulations and 
thermodynamic integration to compute total free energies and perform coexistence analyses.
Nor did we investigate the glass transition,\cite{mckenna-jcp2014} for which purpose 
molecular dynamics simulation would be better suited.
Nevertheless, our results suggest that compressibility of microgels may significantly 
alter the phase diagram of strictly Hertzian spheres.\cite{frenkel2009}

To explore dependence of bulk properties on particle size and pair interactions,
we studied two systems distinguished by widely different particle sizes and interactions.
The first system is a suspension of relatively small microgels (nanogels) of 
dry radius $a_0=10$ nm, comprising $N_m=2\times~10^5$ monomers -- consistent 
with monomers of radius 0.3 nm -- and $N_{\rm ch}=200$ chains, corresponding 
to 0.05 mol \% of cross-linker (assuming two chains per cross-linker).  
Focusing on polymers in a good solvent, e.g., PNIPAM or PAA in water, we took $\chi=0$.  
For these loosely cross-linked particles, the Flory-Rehner free energy in the 
dilute limit [Eq.~(\ref{FloryF})] attains a minimum at swollen radius $a_s=35.055$ nm 
($\alpha=3.5055$).  For particles of this size with Young's modulus $Y=100$ kPa 
and Poisson ratio $\nu=0.5$,\cite{matzelle2003} Eq.~(\ref{epsilon}) yields a 
reduced Hertz pair interaction amplitude $\epsilon^*\simeq~1.5\times~10^3$.
To explore sensitivity to particle elasticity, we also considered microgels 
with $\epsilon^*\simeq~1.5\times~10^4$.  

The second system studied contains much larger particles, of dry radius $a_0=100$ nm,
comprising $N_m=2\times~10^8$ monomers, more closely matching many experimentally 
studied materials.  To again model loosely cross-linked particles in a good solvent, 
we chose $N_{\rm ch}=2\times~10^5$ (0.05 mol \% cross-linker) and $\chi=0$.  
The Flory-Rehner free energy [Eq.~(\ref{FloryF})] is now a minimum at swollen radius 
$a_s=350.55$ nm (dilute limit).  Young's moduli in the range $Y=50-1000$ kPa correspond 
to reduced Hertz pair interaction amplitudes $\epsilon^*\simeq~10^6-10^7$.
We note that the densities and temperatures ($\epsilon^*$ values) studied here 
correspond to thermodynamic states that fall well within the range of fcc crystal 
stability in the known phase diagram of Hertzian spheres.\cite{frenkel2009}  Indeed, 
we never observed structural reassembly from fcc into bcc or any other crystalline structure.

Normalized probability distributions for the equilibrium swelling ratio are shown in 
Figs.~\ref{fig3} and \ref{fig4} over a range of concentrations.  Below close packing 
($\zeta<0.74$), steric (Hertzian) interparticle interactions are sufficiently weak that 
the particles are not significantly compressed, as reflected by the distributions differing negligibly 
from the intrinsic dilute-limit distribution [Eq.~(\ref{FloryP})].  With increasing $\zeta$,
as close packing is approached and exceeded, the distributions not only shift toward smaller $\alpha$, 
as particles become compressed, but also broaden, i.e., become more polydisperse.
The smaller particles (Fig.~\ref{fig3}) have a polydispersity (ratio of standard deviation 
to mean) that increases from roughly 2\% to 4\% as $\zeta$ increases from 0 to 2.  
In contrast, the larger particles (Fig.~\ref{fig4}) have negligible polydispersity 
when uncrowded, but fluctuate significantly in size for $\zeta\ge 1$.  For both 
small and large particles, with increasing Hertz pair potential amplitude, the 
degree of compression {\it and} the breadth of polydispersity increase.
The reason why the polydispersity distribution broadens with increasing concentration
is that compression forces particles into a range of sizes in which the curvature
of the Flory-Rehner free energy [Eq.~(\ref{FloryF})] with respect to $\alpha$ is 
weaker than for uncompressed particles. 
%
Measurements of particle size in suspensions of microgels whose dry radii have a static 
polydisperse\cite{tata2011} or bidisperse\cite{nieves-pnas2016} distribution show that 
swollen particles have an equilibrium polydispersity that decreases with 
increasing particle compression.  Conversely, our study demonstrates that microgels 
whose dry radius is monodisperse have swollen polydispersity -- associated purely with 
fluctuations in swelling ratio -- that increases with particle compression.

Particle compressibility is illustrated further in Fig.~\ref{fig5}, which plots the
volume fraction $\phi$ vs.~generalized volume fraction $\zeta$.  Below close packing,
where particle compression is negligible, $\phi=\zeta$.  With increasing density, 
however, interparticle interactions become stronger and so energetically costly,
relative to the free energy cost of compression, that the particles more readily 
contract to minimize interactions.  As a result, the volume fraction is lower than
the generalized volume fraction ($\phi<\zeta$) for $\zeta>0.74$, the density 
at which nearest-neighbor particles in the fcc crystal would begin to overlap.
For these systems, significant particle compression occurs only in the solid phase.
Moreover, the higher the Hertz pair potential amplitude -- for given values of
$N_m$, $N_{\rm ch}$, and $\chi$ in the Flory-Rehner free energy -- the more easily 
the particles yield to compression.

To explore the evolution of structure as a function of density and interparticle
interactions, and to aid our diagnosis of the liquid-solid phase transition, 
we computed radial distribution functions and static structure factors, as described
in Sec.~\ref{methods}.  Figure~\ref{fig6} shows our results for $g(r)$ over a range
of densities that span the freezing transition.  The emergence of distinct peaks
signals the onset of crystallization at $\zeta\simeq 0.58$ for $a_0=10$ nm and 
$\zeta\simeq 0.49$ for $a_0=100$ nm (black curves in Fig.~\ref{fig6}).  For both 
particle sizes, the freezing density is insensitive to $\epsilon^*$ over the range 
of values considered.  Suspensions of larger microgel particles thus crystallize 
at the same volume fraction as hard-sphere fluid, while suspensions of smaller 
nanogel particles remain fluid up to significantly higher volume fractions.  
This interesting difference in phase stability can be attributed to the relatively 
softer Hertz pair repulsion between the smaller particles.

Figure~\ref{fig7} shows corresponding results for $S(q)$, computed using Eq.~(\ref{Sq}) 
by averaging over the same configurations used to compute $g(r)$ in Fig.~\ref{fig6}.  
With increasing density, the peaks progressively sharpen and shift toward larger $q$.  
At the densities at which distinct peaks in $g(r)$ indicate crystallization, 
the main peak of $S(q)$ attains a height in the range $S(q_{\rm max})=2.8-3$.
This freezing threshold is consistent with the Hansen-Verlet criterion for fluids 
interacting via a Lennard-Jones pair potential,\cite{hansen-verlet1969} according to 
which $S(q_{\rm max})\simeq 2.85$ at freezing.  In passing, we note that a different 
freezing criterion proposed specifically for ultrasoft pair-potential 
fluids\cite{likos-loewen2001} does not apply here, since the Hertz pair potential, 
although bounded, has amplitudes in our system that far exceed the thermal energy $k_BT$.
As a consistency check, we also computed the Lindemann parameter $L$, defined as 
the ratio of the root-mean-square displacement of particles from their equilibrium 
lattice sites to the lattice constant, and confirmed that $L<0.15$ and remains 
constant in the solid phase.

Finally, we turn to the osmotic pressure, which we computed from our simulations using
Eq.~(\ref{pressure}).  Figure~\ref{fig8} presents plots of osmotic pressure vs.~generalized 
volume fraction (equation of state) for the two microgel systems studied.  In the dilute regime, 
the osmotic pressure is very close to that of the hard-sphere fluid, which is consistent 
with the swelling ratio results (Fig.~\ref{fig5}).  At volume fractions above 20\%, 
deviations emerge and steadily grow with increasing volume fraction.  The compressible 
microgels have osmotic pressures consistently lower than for the hard-sphere fluid.
At volume fractions approaching close packing, the microgel osmotic pressure rapidly rises,
yet remains finite beyond close packing, in sharp contrast to the hard-sphere fluid, 
whose pressure diverges at close-packing.  Furthermore, as is clear from Fig.~\ref{fig8}a, 
the osmotic pressure rises more gradually with concentration the smaller and softer 
the particles.  This equilibrium trend may have implications for rheological properties,
such as shear viscosity.  In fact, recent measurements\cite{dutcher-preprint2016} of 
the zero-shear viscosity of aqueous suspensions of monodisperse, highly branched, 
phytoglycogen nanoparticles\cite{dutcher-biomacromol2016} are consistent with 
significant compression of the particles.

At the apparent freezing density of the microgel suspensions, as diagnosed by the 
main peak height of the static structure factor, the osmotic pressure plateaus, 
consistent with a thermodynamic phase transition between fluid and solid.  
Thus, the freezing behavior of the compressible microgels modeled here seem to be
accurately described by the Hansen-Verlet criterion.  Our methods are not suited, 
however, to identifying a glass transition, which could preempt crystallization.

Finally, to probe the dependence of system properties on solvent quality, we have
repeated several of the calculations for $\chi=0.5$, modeling a theta solvent.
As $\chi$ increases, the particles become progressively compressed, as would be expected 
with worsening solvent quality,  and exert lower osmotic pressure.  However, with 
varying concentration, the systems display similar qualitative trends in polydispersity,
volume fraction, structure, and osmotic pressure.

\section{Conclusions}\label{conclusions}
In summary, we designed and implemented Monte Carlo computer simulations of model suspensions of 
compressible, soft, spherical, colloidal particles that fluctuate in size and interact via an 
elastic Hertz pair potential.  For this purpose, we introduced a novel trial move that allows 
particles to expand and contract, according to the Flory-Rehner free energy of cross-linked 
polymer networks, in response to interactions with neighboring particles.  From a series of 
simulations over ranges of particle parameters chosen to be consistent with experimental systems, 
we analyzed the dependence of interparticle correlations and thermodynamic behavior on 
particle size (dry radius) and softness (elastic modulus) by computing a variety of 
thermal and structural properties.  

Radial distribution functions, static structure factors, and osmotic pressures all display 
behavior similar to that of a hard-sphere fluid at low densities -- volume fractions below 
about 0.3 -- but reveal the particles' intrinsically soft nature at higher densities.
Swelling ratios confirm that, with increasing density above close packing, the particles 
become progressively compressed and polydisperse, compared with their dilute (uncrowded) states.
While suspensions of relatively large, stiff particles (microgels of dry radius 100 nm) 
crystallize at the same volume fraction as a fluid of hard spheres, internal degrees of 
freedom associated with particle compressibility and size polydispersity enable suspensions 
of smaller, softer particles (nanogels of dry radius 10 nm) to remain in a stable fluid phase 
up to densities significantly beyond freezing of the hard-sphere fluid.  For both nanogels 
and microgels, the freezing transition is accurately predicted by the Hansen-Verlet criterion.

Our calculations of equilibrium swelling ratios and bulk osmotic pressures are qualitatively consistent 
with observations of significant compression of deformable particles, albeit ionic microgels, 
only at volume fractions approaching and exceeding close packing.\cite{weitz-jcp2012,nieves-prl2015}  
Similarly, our finding that particle softness lowers the osmotic pressure is consistent 
with experimental measurements of reduced zero-shear viscosity in suspensions of soft 
nanoparticles.\cite{dutcher-preprint2016} Furthermore, our calculations may guide the choice 
of system parameters in future experiments and molecular-scale simulations and could be extended 
to map out equilibrium phase diagrams.

The coarse-grained model of nonionic, spherical microgels studied here may be 
refined by incorporating a more realistic swelling free energy\cite{potemkin2015} 
that goes beyond the limitations of the Flory-Rehner theory noted in Sec.~\ref{particles},
and may be further developed in several directions.  
For example, the model may be extended to ionic microgels, whose properties 
emerge from an interplay between elastic and electrostatic interparticle interactions.
Such an extension would enable analyzing the coupled influences of fluctuating particle size 
and counterion distribution on swelling, structure, and phase behavior of concentrated 
suspensions of soft, charged colloids.  By incorporating shape fluctuations, our simulation 
method also may be generalized to model dynamically deformable particles, as in recent studies 
of elastic colloids,\cite{riest2015,cloitre-bonnecaze2010,cloitre-bonnacaze2011} and 
extended to study the influence of added hard crowders, as in our previous studies 
of polymer-nanoparticle mixtures.\cite{lim-denton2014,lim-denton-JCP2016,lim-denton-SM2016}
Finally, rheological properties, such as diffusion coefficients, viscosity, elastic moduli,
jamming and glass transition densities\cite{mckenna-jcp2014} could be explored by wedding 
our model of soft, compressible, size-fluctuating colloids to appropriate simulation methods,
such as Brownian dynamics, lattice Boltzmann, or molecular dynamics of 
micromechanical models.\cite{cloitre-bonnecaze2010,cloitre-bonnacaze2011}

\vspace*{0.5cm}
\noindent{\bf \large Acknowledgments} \\[1ex]
This work was partially supported by the National Science Foundation (Grant No.~DMR-1106331).

\balance

\bibliographystyle{rsc}

\providecommand*{\mcitethebibliography}{\thebibliography}
\csname @ifundefined\endcsname{endmcitethebibliography}
{\let\endmcitethebibliography\endthebibliography}{}

\end{document}